\documentclass{ws-ijmpd}
\usepackage[super,compress]{cite}
\usepackage{psfig}
\begin{document}

\markboth{Popov, Postnov, Pshirkov}
{Fast radio bursts: superpulsars, magnetars, or something else?}

%
\catchline{}{}{}{}{}
%

\title{FAST RADIO BURSTS: SUPERPULSARS, MAGNETARS, OR SOMETHING ELSE?  }

\author{SERGEI POPOV}

\address{Sternberg Astronomical Institute, Lomonosov Moscow State
University, Universitetskii pr. 13\\
Moscow 119234, Russia\\
polar@sai.msu.ru}

\author{KONSTANTIN POSTNOV}

\address{Sternberg Astronomical Institute, Lomonosov Moscow State
University, Universitetskii pr. 13\\
Moscow 119234, Russia\\
kpostnov@gmail.com}

\author{MAXIM PSHIRKOV}

\address{Sternberg Astronomical Institute, Lomonosov Moscow State
University, Universitetskii pr. 13\\
Moscow 119234, Russia\\
pshirkov@sai.msu.ru}

\maketitle

\begin{history}
\received{Day Month Year}
\revised{Day Month Year}
\end{history}

\begin{abstract}
We briefly review main observational properties of fast radio bursts (FRBs) and discuss two most popular hypothesis for the explanation of these enigmatic intense millisecond radio flashes. FRBs most probably originate on extragalactic distances, and their rate on the sky is about a few thousand per day with fluences above $\sim$~1~Jy~ms (or with fluxes larger than few tenths of Jy). Two leading scenarios describing these events include strong flares of magnetars and supergiant pulses of young radio pulsars with large rotational energy losses, correspondingly. At the moment, it is impossible to choose between these models. However, new telescopes can help to solve the puzzle of FRBs in near future.
\end{abstract}


\ccode{PACS numbers:  	95.85.Bh, 95.85.Fm, 97.60.Jd}


\section{Introduction}	

Fast radio bursts (FRBs) are intense radio flares with millisecond duration demonstrating large peak fluxes ranging from a few hundredths of Jy up to $> 100$ Jy (see a recent review in Ref.~\refcite{rev17}). All known FRBs are characterized by large dispersion measures (DM) $\sim 200-3000$ pc cm$^{-3}$, which cannot be explained by the interstellar medium in the Milky Way galaxy. The first FRB010724 has been reported 10 years ago --- in 2007 \cite{lor}. However, active FRB studies began only in the second half of 2013 when four other similar events have been discovered \cite{thor} bringing the total number of bursts to six. The high DM suggests extragalactic distances ($\sim$ Gpc), so the bright peak flux and short millisecond duration imply a very high brightness temperature of the observed radio emission (about  $10^{35}-10^{36}$~K) and hence a non-thermal radiation mechanism. 

The occurrence rate of FRBs estimated from observations is very high: several thousand events a day over the sky (it is not clear, yet, how much this number can be increased due to dimmer bursts with peak fluxes $\lesssim$~few~$\times 0.1$~Jy or/and fluences $\lesssim$~1~Jy~ms), much exceeding that of many other known transients (e.g., gamma-ray bursts or coalescences of compact objects), but significantly smaller than the rate of supernovae (note, that for all FRBs detected in real-time no accompanying transients have been detected at any energy range), see Ref.~\refcite{law17}. For extragalactic FRBs the estimated rate corresponds to $\sim$ 100 events a day from the local $\sim$~1~Gpc$^3$ volume \cite{lor}. However, due to a small field of view of the existing radio telescopes, only a small fraction of the bursts is detected.

\begin{figure}[pb]
\centerline{\psfig{file=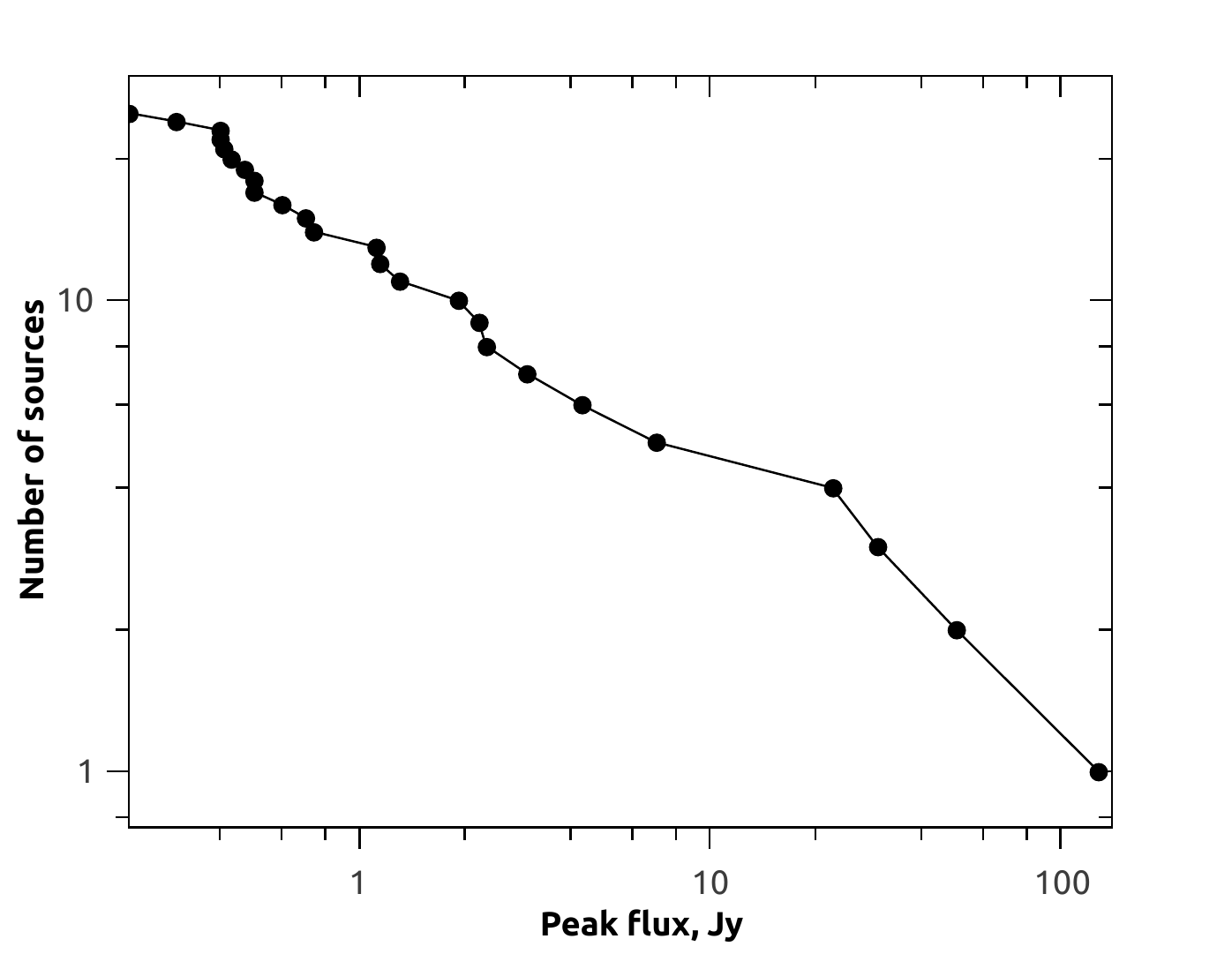,width=13cm}}
\vspace*{8pt}
\caption{Log N -- Log S distribution for fast radio bursts. Data from http://frbcat.org have been used.
\label{f1}}
\end{figure}

Presently, about 30 FRBs have been recorded (see the on-line catalog on http://frbcat.org, description is provided in Ref.~\refcite{catalogue}).  Log N -- Log S distribution for known event is shown in Fig.\ref{f1}.  Most of them ($>70\%$) was discovered by the 64-m Parkes radio telescope in Australia. Mostly FRBs are detected at 1.4 GHz. All searches for emission below $\sim 800$~MHz produced zero results (for example, LOFAR did not detect any FRBs, see Ref.~\refcite{lofar}). Generally, spectra of FRBs are not well-constrained \cite{bs16}. Different bursts of the only repeating source demonstrate significant spectral variations \cite{spit16}.

\begin{figure}[pb]
\centerline{\psfig{file=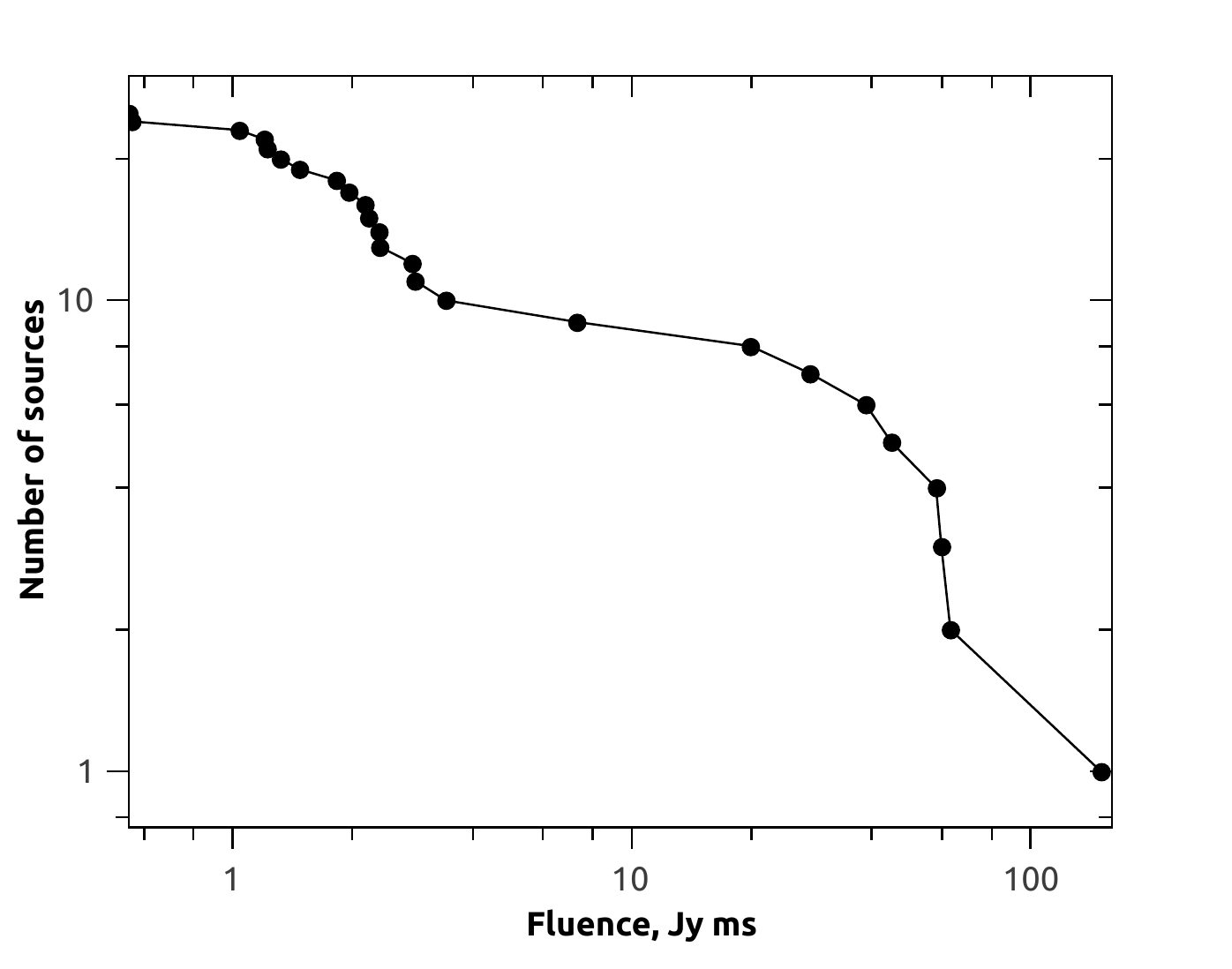,width=13cm}}
\vspace*{8pt}
\caption{Log N -- Log F distribution for fast radio bursts. Data from http://frbcat.org have been used.
\label{f3}}
\end{figure}

\begin{figure}[pb]
\centerline{\psfig{file=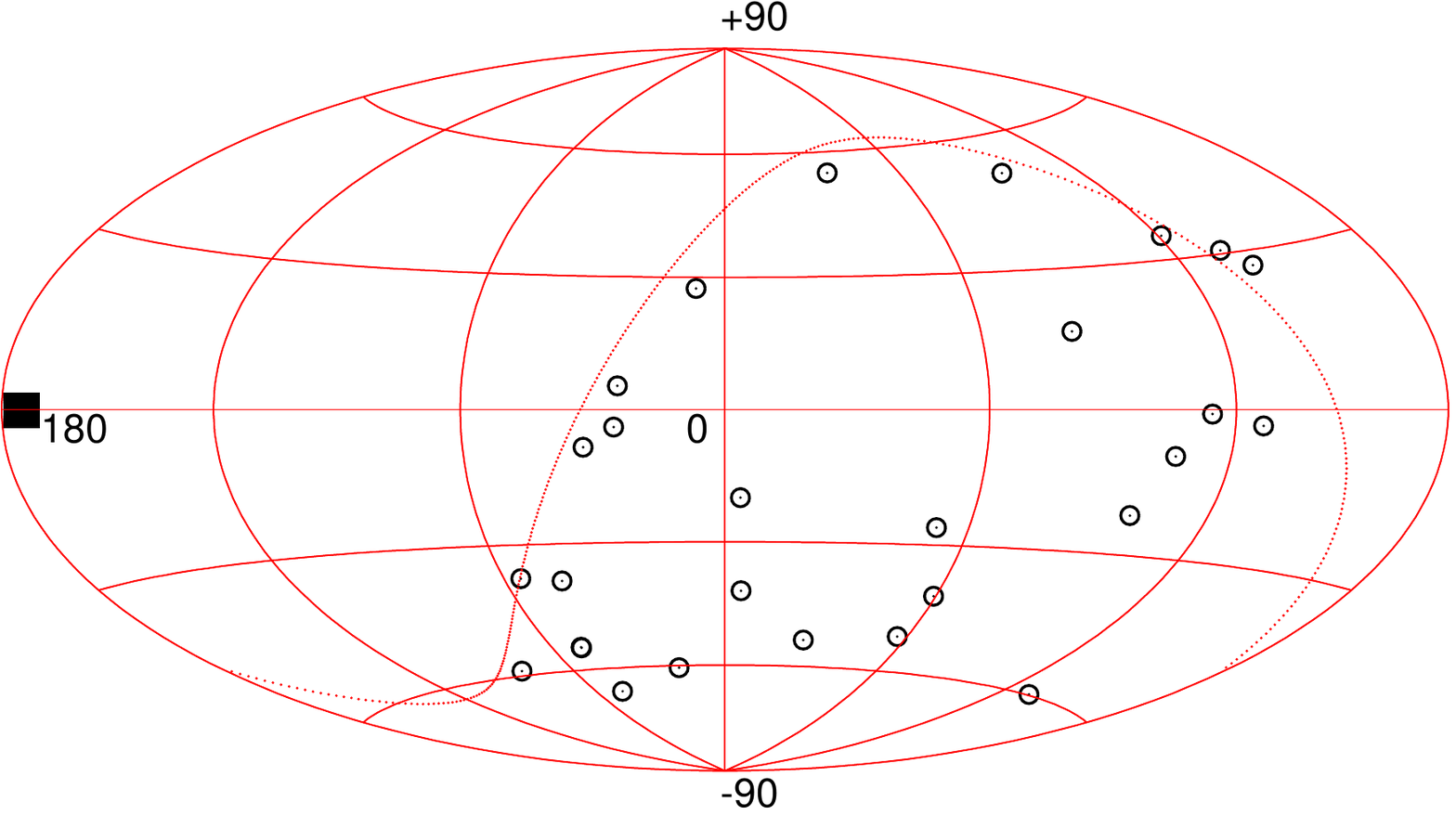,width=13cm}}
\vspace*{8pt}
\caption{Distribution of FRBs in the sky in Galactic coordinates. Circles show non-repeating FRBs. Most of them are detected at Parkes or by UTMOST, both in Australia. Square shows the repeating source detected at Arecibo. Dotted line -- celestial equator. Data from http://frbcat.org have been used.
\label{f2}}
\end{figure}

The repeater --- FRB121102, --- is found to produce bursts without any detected periodicity \cite{spit14}. More than few hundred events are already detected from this source. For this object the host galaxy was identified \cite{host}. For other sources no repetitions of activity are found, despite intensive searches.  However, in the case of FRB121102 repeating bursts are on average weaker then typical flares from other FRBs \cite{spit16}. Thus, in some cases repeating activity of other FRBs can avoid detection with instruments much smaller than the 300-meter telescope in Arecibo, where FRB121102 (and majority of its repeating bursts) has been discovered.

Integral distribution in fluence (Log N -- Log F) is shown in Fig.\ref{f3}. Here fluence was calculated using the data from on-line catalogue as $Fluence=Peak flux \times Width$. The distribution has a rather peculiar form. It looks bimodal, and each mode can be fitted with a linear distribution in the linear scale (this was first noted in Ref.~\refcite{lbp16}). However, this feature can be an effect of incomplete observational data for dim sources.

All available data are consistent with cosmological origin of these events \cite{li17}.  Sky distribution of known sources, see Fig.~\ref{f2}, does not look uniform because most of FRBs have been detected by telescopes in Australia (Parkes, UTMOST) in specific surveys with non-uniform sky coverage.\footnote{New projects -- CHIME, Apertif, -- might help to identify many FRBs in the northern sky.} However, detailed analyses indicates that all available data are in correspondence with uniform sky distribution which is expected in case of  extragalactic origin of FRBs at distances larger than $\sim 100$~Mpc. 

The nature of FRBs remains unknown. About twenty different hypothesis have been proposed (and with variations their number reaches a few dozens!). Some of them already can be considered as being ``refuted'' (for example, because of too low rate or due to undetection of any transient counterparts of FRBs), some look rather exotic (like models related to cosmic strings or charged black holes). Most conservative approaches link FRBs to some kind of activity of young neutron stars. Presently, the two leading approaches relate these millisecond bursts of radio emission to the activity of magnetars or to very strong pulses of energetic radio pulsars. Below we briefly describe both models and then present our conclusions.

FRBs could also serve as valuable probes of extragalactic medium and instruments to study a wide class of models of fundamental physics. Simultaneous measurement of DMs, RMs,  and distances to the bursts will allow to robustly  map gas and magnetic fields in the Large Scale Structure \cite{2016ApJ...824..105A}. If these bursts occur at high redshifts, $z>1$, they can serve as a very sensitive tool for the precision cosmology, especially for measurements of the baryonic  content of the Universe  \cite{2014ApJ...783L..35D}. Global cosmological parameters can also be  estimated using observations of  FRBs \cite{2014PhRvD..89j7303Z}. In any case, these methods need large quantity of observed FRBs ($\mathcal{O}(1000)$) and will be fully implemented only in the era of the SKA.

Very short duration of the FRBs makes them one of the most efficient ways to test the  validity of the equivalence principle. If this principle is violated,  parts of the signal at different frequencies would  have a different propagation times in the gravitational field of the Galaxy  \cite{2015PhRvL.115z1101W}. Also the most stringent limits on the mass of the photon were obtained using the robust identification of the host galaxy of the FRB121102 and distance to it \cite{2017PhLB..768..326B}: $m_\gamma<2.2 \times 10^{-14}$~eV, i.e. $<3.9\times 10^{-47}$~g.

\section{Magnetar model}

Already in 2007, immediately after the first FRB was reported, the paper \refcite{pp07} suggested that such short radio bursts can be related to hyperflares of magnetars -- neutron stars with very strong magnetic fields which can be rapidly released due to reconfiguration of the outer field structure (see a review of magnetars in Ref.~\refcite{tzw}). This scenario seems very plausible, as statistical (the rate of hyperflares is once in at least $\sim $~few hundred years per galaxy, or even more seldom, see  Ref.~\refcite{pp06,laz06}) and energetic considerations allow one to explain basic properties of FRBs with the transformation of just a tiny fraction ($\sim 10^{-5}$~--~$10^{-6}$) of the magnetar giant or hyper flare energy ($\sim 10^{44}-10^{46}$ erg) into a short radio burst ($\sim 10^{39}-10^{41}$ erg). Also the time scale of two types of transients match well as intense magnetar bursts have very sharp rising fronts\footnote{In addition, in Ref.~\refcite{pp07} it was mentioned that the birth rate of magnetars is compatible with the rate of FRBs.}.

In addition to statistical and energy properties of FRBs, the magnetar model can easily explain absence of counterparts in other wavelengths. If a magnetar flare happens at a distance significantly larger than few tens of Mpc, then it is impossible to detected the burst at high energy range with present day $\gamma$-ray monitors, see Ref.~\refcite{pp06} and references therein. Thus, the FRB might not be accompanied by a gamma-ray burst or any kind of afterglow directly related to the emission of the magnetar flare, in accordance with observations \cite{pal14, yam16, xi17}. 

FRBs from magnetars, if they are produced in hyperflares or similar events, might repeat, but the time scale for this repetition is very long --- at least hundreds of years, except, may be, a short period in the very youth of a neutron star, which may explain the repeating source of FRBs. Sources might be situated, on average, in regions of intensive starformation \cite{pp07, pp06}, as magnetars are mostly young neutron stars with ages $\lesssim$~few tens of thousand years.  

A working model of the radiation mechanism for FRBs from magnetar flares was elaborated by Yuri Lyubarsky \cite{l14}. After that, several modifications of this scenario were proposed, see for example Ref.~\refcite{mkm}.
In this scenario, the coherent radio emission is generated by the synchrotron maser mechanism operating at a relativistic shock. The forward shock is produced when a strong electromagnetic pulse from the magnetar hyperflare propagates from the magnetar's magnetosphere outwards and meets a density discontinuity (for example, the boundary between magnetar's pulsar-wind nebula and the ambient medium). Note that pulsar-wind nebulae have been indeed found around some Galactic magnetars \cite{Vink_ea09}.  On other hand, not necessarily every bursting magnetar is a source of FRB, as not all of them are situated in proper surroundings, from the point of view of the model proposed in Ref.~\refcite{l14}.

An important prediction of this scenario is a strong TeV burst arising simultaneously with the radio pulse due to synchrotron radiation of electrons at the forward relativistic shock. Observations by ground-based gamma-ray telescopes (such as VERITAS and H.E.S.S., and CTA in future) can prove or falsify this prediction. 

\section{Superpulses by radio pulsars}

The idea that FRBs can be analogues of giant pulses of energetic radio pulsars was proposed in 2015 \cite{cw}\footnote{See, however, a note in sec.~1 of Ref.~\refcite{pp07} where it was mentioned that scaling of giant pulses of the Crab pulsar with $\dot E \sim \mu^2 P^{-4}$, where $\mu$ is magnetic moment and $P$ -- spin period, results in radio fluxes compatible with a FRB from even $\sim 500$ Mpc distance.} Indeed, scaling of the most powerful ``shots'' observed in the Crab pulsar to a neutron star with millisecond spin period and magnetic field above $\sim 10^{13}$~G (but not necessarily with magnetar-scale fields) results in bursts potentially similar to FRBs if they are situated at distances $\sim $~few hundred Mpc (we underline here that the precise distance scale for FRBs is still unknown). Giant pulses of the Crab pulsar are known to be very short (below one millisecond), so similar events can fit timing properties of FRBs, as width of signals from many of these sources can be dominated by propagation effects, i.e. not by intrinsic properties. 

This scenario was later discussed and developed in many papers, see for example Ref.~\refcite{c16,p16} and references therein. In the article Ref.~\refcite{lbp16} the authors studied evolution of the DM due to a young supernova remnant around the pulsar. This can explain large total DM even if distances are significantly below Gpc scale. In this model, any detected repeating burst will display a DM quickly decreasing with a characteristic  time of  several years as  young supernova remnants evolve on this scale. This prediction can be tested soon, if consequent bursts from the same sources are detected (it is expected that in this model each source could produce new pulses every few days, see Ref.~\refcite{lbp16} for details. This rate of repetition is still possible for most of the known sources). Also, this model naturally explains the lack of FRB detections at low frequencies -- the free-free absorption in the dense medium of the remnant precludes radiation at frequencies lower than $700-800~$MHz from leaving the shell.  

The model of supergiant bursts leads to an important prediction: as the FRB sources are ``local'' in some sense, $D\lesssim 100-200$~Mpc, one would expect some degree of correlation with the positions of galaxies in this volume. Large localization errors and small number of detected bursts do not allow us to robustly check this prediction at the moment, but it is expected that observation of $\mathcal{O}(100)$ sources will be sufficient for this purpose. Also, these young neutron stars   might be strong X-ray sources due to huge rotational energy losses, as it is well-known in the case of Galactic energetic radio pulsars \cite{pp16}. This prediction will also be checked with better  localization of the FRBs. The increasing statistics of FRBs and rapid localization within small error boxes can be done soon (may be even within 1-2 years) with new radio astronomical facilities.


\section{Conclusions}

After 10 years of exploration, fast radio bursts remain a major puzzle. Despite new important discoveries (real-time detections with intensive follow-up at different wavelengths, detection of circular and linear polarization, the detection of one repeating source and identification of its host galaxy -- see a recent review in Ref.~\refcite{rl17}), it is unclear whether all these sources represent a single population, what is their exact distance scale, and hence their typical luminosity. 

Most likely, the engine of these enigmatic bursts is related to neutron stars. Either the magnetic (in the magnetar scenario) or rotational (in the case of supergiant pulses) energy of a neutron star is transformed into a strong radio flare with enormously high brightness temperature up to $\sim 10^{36}$~K. General properties of FRBs can be reproduced in any of these models. Both models can also explain the repeating FRB 121102 making it just a peculiar source (may be at a specific evolutionary stage, for example, a very young neutron star with significant rotational energy losses and/or frequent and violent reorganization of the electric currents supporting the magnetosphere) of generally the same nature. Interestingly, properties of the host galaxy of the repeating source are consistent with both models. The galaxy demonstrates high star formation rate, so the appearance of a young neutron star (a magnetar, or a very energetic pulsar) seems to be quite natural \cite{host}. 

However, some important questions remain unanswered. For example, why up to now, despite intensive searches, FRB pulses have not been detected at frequencies below $\sim 600$~MHz?  What mainly contributes to the observed DM: the intergalactic medium, or immediate surroundings of the source (for example, a dense supernova remnant shell or may be interstellar matter in a star formation region)? Do bursts from most of objects repeat on some longer time scale (starting from $\sim$~few days, as repetition on this time scale is not completely ruled out), or repeaters (of which we know a single one, yet) form a completely separate class of sources? Hopefully, in the near future, the increasing FRB statistics which will be obtained with new observational facilities (FAST, Apertif, UTMOST, CHIME, HIREX, MeerKAT, etc.) can help to solve this puzzle. Then FRBs can be used as probes to study intergalactic medium,  may be as a kind of standard candles in cosmology, and, finally, as a tool to test fundamental theories.

\section*{Acknowledgments}
This work has been supported by the Russian Science Foundation grant 14-12-00146.
S.B.P. is grateful to M. Lyutikov for interesting discussions on FRBs.

\end{document}